# EquiSay, Órtesis antiequina con sensor de electromiografía.

M. Terradillos Perea[1], O. Alonso González[1], C. Soguero Ruiz[1], D. Gutiérrez[2]

[1]Escuela de Ingeniería de Fuenlabrada, Universidad Rey Juan Carlos, Madrid, España

m.terrradillos.2021@alumnos.urjc.es, o.alonso.2020@alumnos.urjc.es, cristina.soguero@urjc.es

[2]Departamento de Fisioterapia, Hospital Universitario de Móstoles, Móstoles, España

david.gutierrez@salud.madrid.org

## Resumen

*El pie equino es una alteración neuromuscular que afecta la dorsiflexión del pie dificultando la marcha y comprometiendo la calidad de vida de quienes lo padecen [3]. El siguiente estudio presenta EquiSay, una ortesis antiequina dinámica equipada con un tensor elástico anterior y un sensor electromiográfico (EMG) capaz de medir la acción y evolución de los músculos implicados en la rehabilitación del pie equino (tibial anterior principalmente) [6]. EquiSay se presenta como una solución capaz de proporcionar un soporte dinámico facilitando la corrección de la postura del pie y mejorando el movimiento natural. Además, la incorporación del sensor EMG al diseño permite registrar la actividad muscular en tiempo real brindando información valiosa para los profesionales de la salud. Tras el diseño y aplicación práctica del dispositivo en cuestión, en el presente estudio se mostraron resultados positivos mostrándose los pacientes satisfechos y resultando para los clínicos una gran ayuda para ajustar tratamientos de manera más precisa gracias a la capacidad de monitorear la respuesta neuromuscular del paciente en tiempo real [4].*

## 1. Introducción y contexto

El desarrollo de dispositivos de asistencia para la rehabilitación motora ha avanzado significativamente en los últimos años permitiendo mejorar la calidad de vida de aquellas personas que padecen trastornos neuromusculares [5]. Dentro de estas afecciones, el pie equino representa una limitación funcional común que dificulta la marcha debido a la incapacidad de realizar la dorsiflexión del pie. Estos pacientes únicamente apoyan la zona anterior del pie sin plantar el talón ("andan de puntillas") y alterando como resultado el patrón normal de la marcha, siendo esta inestable y generando por tanto mayor riesgo de caídas [3]. Esta condición es común en enfermedades neurológicas como el accidente cerebrovascular, la parálisis cerebral o lesiones medulares. También puede ser congénita como causa de una malformación durante el desarrollo del feto (el caso de la espina bífida), traumática o relacionada con procesos reumáticos [1].

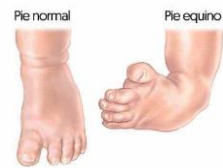

***Figura 1.*** *Comparativa pie equino varo de causa congénita con pie normal infantil.*

### 1.1. Musculatura Implicada

Como se explicaba anteriormente, el pie equino es una deformidad caracterizada por una flexión permanente del pie hacia abajo, lo que hace que el talón no toque el suelo al caminar y como consecuencia afecte a varios músculos de la pierna y el pie.

En cuanto a músculos flexores plantares, el tríceps sural es el principal responsable de la flexión plantar del pie y está compuesto por los músculos gastrocnemio y sóleo. El gastrocnemio permite movimientos como ponerse de puntillas o empujar el pie contra el suelo al caminar y es esencial en la fase de propulsión de la marcha, ayudando a levantar el talón del suelo. En aquellos pacientes con pie equino, el gastrocnemio suele estar acortado o espástico, lo que genera una sobre activación de la flexión plantar y dificulta la dorsiflexión del pie [6]. El soleo se encuentra debajo del gastrocnemio y cuando el pie está en posición de equino también se acorta y se tensa, lo que puede provocar dolor y rigidez en la pierna.

El tibial anterior (como musculo principal) y los extensores largos de los dedos (musculatura accesoria) son músculos participantes en la dorsiflexión del pie. Una condición de debilidad o parálisis en el tibial anterior puede favorecer la posición en equino. Este musculo situado en la parte medial de la pierna, se encarga de elevar el pie hacia arriba permitiendo que los dedos no arrastren durante la fase de oscilación de la marcha. Además, regula el descenso controlado del pie tras el contacto inicial con el suelo evitando que el pie caiga bruscamente. Por tanto, la falta de activación de este musculo deriva en una marcha en estepaje y contracturas de los flexores plantares.

Finalmente, como músculos eversores involucrados están los peroneos largo y corto. Estos músculos facilitan la eversión del pie y aportan estabilidad. Su debilidad puede contribuir a la deformidad en equino varo. Cuando el pie está en posición de equino, los músculos peroneos se estiran y debilitan, esto genera una inversión excesiva del pie y una marcha inestable. Cuando se da el caso contrario, los músculos peroneos están hiperactivos y los tibiales debilitados y se produce una eversión excesiva del pie y como consecuencia pie equino valgo.

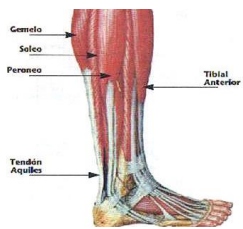

*Figura 2.Musculatura implicada. (Representados: tríceps sural, músculo tibial anterior, peroneos y tendón de Aquiles).*

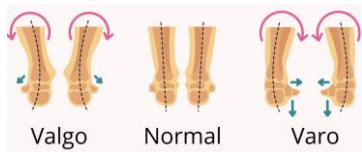

*Figura 3.Comparativa y proceso de pies valgo, normal y varo.*

## 1.2. Tratamientos

El tratamiento del pie equino dependerá de la causa, gravedad de la deformidad y edad del paciente pudiendo ser conservador o quirúrgico [1].

Como posibles tratamientos conservadores en la actualidad existen diferentes técnicas como la fisioterapia, los ejercicios de fortalecimiento y estiramiento, el uso de ortesis o aparatos ortopédicos, y la terapia ocupacional.

La fisioterapia y rehabilitación son técnicas centradas en mejorar la flexibilidad y fuerza de los músculos de la zona afectada para combatir el efecto del pie caído, corregir la postura del pie y en último caso mejorar la funcionalidad del pie. Especial importancia de movimientos de dorsiflexión y uso combinado de férulas antiequinas. El empleo de aparatos ortopédicos, como férulas o soportes de tobillo-pie, ayuda a mantener la posición adecuada del pie y facilitan la marcha ya que ayudan a levantar la punta del pie, evitando su arrastre y manteniendo una posición más horizontal. De esta manera, se compensa la debilidad de los músculos extensores oponiendo una resistencia a la flexión plantar en el contacto del talón y durante la fase de balanceo de la marcha. Destaca el método de Ponseti. Método de referencia utilizado para los casos de pie equino varo congénito idiopático [2]. Aquí el fisio manipula suavemente el pie y coloca yesos semanales para ir corrigiendo gradualmente la posición del pie y una vez lograda la alineación deseada, se utiliza una férula para mantener la corrección.

Si la gravedad del problema es superior o los tratamientos conservadores no efectivos, se opta por un tratamiento quirúrgico [1]. Al tratarse de una técnica invasiva, no es la opción más habitual, pero en algunos casos necesaria para sanar el nervio peroneo, alargar tendones, liberar tejidos blandos o realinear estructuras óseas. Existen diferentes técnicas quirúrgicas que se utilizan en la actualidad para corregir el pie equino, como la tenotomía, la osteotomía y la artrodesis y su elección depende de la edad del paciente y la gravedad de la deformidad entre otros.

Es fundamental que el tratamiento del pie equino sea personalizado, considerando la causa específica, la edad del paciente y otros factores individuales, para lograr los mejores resultados posibles.

## 1.3. Ortesis en la actualidad

El presente estudio se ha centrado en el tratamiento del pie equino mediante ortesis antiequinas [4]. Existen varios tipos diseñadas para adaptarse a las diferentes necesidades y condiciones de cada paciente y ayudando asi a mantener el pie en una posición adecuada permitiendo una marcha más natural.

Según su funcionamiento y grado de asistencia que prevean, podemos distinguir entre ortesis pasivas, como las férulas nocturnas de tobillo-pie; activas o dinámicas como las ortesis de tobillo-pie con tensor elástico; y ortesis funcionales con asistencia neuromuscular que utilizan electroestimulación funcional (FES) como Walkaide o Ness L300 [6]. También resulta de interés la clasificación basada en el diseño estructural:

**Órtesis de tobillo-pie (AFO, Ankle-Foot Orthosis):** Las más comunes en este tipo de tratamiento y se diferencia entre ortesis rígidas, articuladas y con resorte posterior.Se extienden desde la parte inferior de la pierna hasta la planta del pie y suelen estar hechas de termoplástico, fibra de carbono u otros materiales ligeros y duraderos

**Órtesis tipo DAFO (Dynamic Ankle-Foot Orthosis):** Ofrecen un soporte dinámico. Ortesis de interés, detallada más adelante por incluirse Equisay en este grupo.

**Férulas nocturnas antiequinas:** Diseñadas para que el paciente duerma manteniendo una posición neutral del pie evitando contracturas en el tendón de Aquiles.

**Órtesis de sujeción con tensor elástico:** Diseños que incluyen bandas elásticas que favorecen la dorsiflexión durante la marcha sin restringir completamente el movimiento.

**Ortesis a medida:** Fabricadas a medida para corregir la alineación del pie y mejoran la distribución del peso en el pie y tobillo. Proporcionan un ajuste preciso para adaptarse a las características anatómicas específicas de cada paciente.

La ortesis ¨EquiSay¨ diseñada se engloba en las ortesis dinámicas (DAFO) con tensor elástico. En los apartados siguientes aparecen detalladas sus especificaciones.

*Figura 4.Ortesis antiequina con tensor elástico (Boxia Plus)*

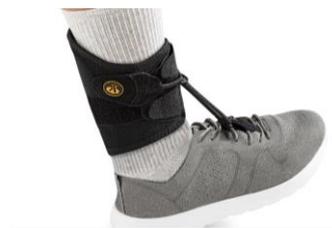

## 2. Metodología

### 2.1 Materiales

Para el diseño de Equisay, la nueva ortesis antiequina, se ha utilizado como base el dispositivo de ayuda externa Cyclalon Db3 de AXZ INNOVATIONS S.L. y un sensor EMG capaz de medir la estimulación de los músculos implicados en la patología estudiada.

EL envío y transferencia de datos registrados se ha hecho posible gracias a un dispositivo Arduino UNO junto a un módulo WIFI incorporado en este. Finalmente, para el ajuste y acople del estimulador a la ortesis, se ha diseñado un sistema de velcros gracias a los cuales ha sido posible adherir una bolsa para el trasporte y sostén del estimulador y la placa mencionada. Esta placa se encuentra encerrada a su vez en una carcasa protectora de plástico dado que es un objeto altamente frágil y sensible. Diseño pensado para ser lo más ligera y cómoda posible, fácilmente ajustable y altamente funcional.

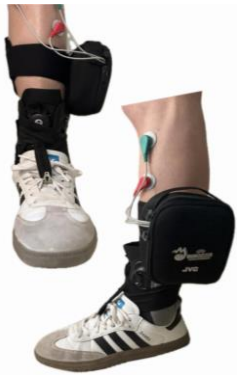

**Figura 5.** Equisey en sujeto de prueba 1

### 2.2 Adquisición y Transferencia de Datos EMG

#### 2.2.1 Comunicación entre el Arduino y el Servidor

El proceso de adquisición de datos comienza con la medición de las señales EMG, las cuales son recogidas por el sensor EMG conectado al microcontrolador Arduino UNO. El código del microcontrolador permite leer las señales generadas por la activación de los músculos y transmitir esta información al servidor en tiempo real. Para ello, se utiliza el protocolo UDP (User Datagram Protocol), que es ideal para la transferencia de datos en tiempo real debido a su baja latencia y la posibilidad de enviar paquetes de datos de manera continua, aunque sin la garantía de entrega de otros protocolos como TCP [8].

Los datos son recibidos por un servidor, implementado en un ordenador con Python. Este recibe los datos y los almacena en un archivo CSV para su posterior análisis. Este archivo permite que los datos sean fácilmente accesibles para su procesamiento y visualización. Además, el servidor es capaz de mostrar las señales en tiempo real, permitiendo al usuario monitorear las condiciones del paciente de manera continua.

#### 2.2.2 Almacenamiento de Datos

Los datos de las señales EMG recibidos desde el Arduino se almacenan en archivos CSV. Cada entrada en el archivo contiene dos componentes: un valor de la señal EMG y una marca temporal que indica cuándo se registra esa señal [10]. Este formato permite tener un registro preciso de la actividad muscular de cada paciente, facilitando su análisis posterior.

### 2.3 Procesamiento de Señales EMG

#### 2.3.1 Preprocesamiento de los Datos

Una vez almacenados los datos de las señales EMG, es necesario realizar un preprocesamiento antes de que puedan ser utilizados para análisis más profundos o alimentados a modelos de inteligencia artificial. Este preprocesamiento incluye varias etapas clave:

**Segmentación de las señales**: Debido a la naturaleza continua de las señales EMG, estas se dividen en segmentos más pequeños. Se deben muestrear las señales a un rango de entre 1 kHz y 2 kHz; sin embargo, esto no es posible por las limitadas capacidades del sensor disponible.

**Filtrado**: Se aplican filtros para eliminar el ruido y los artefactos presentes en las señales EMG. Esto incluye el uso de filtros de paso banda para eliminar las frecuencias no deseadas que podrían interferir con la precisión del análisis [11].

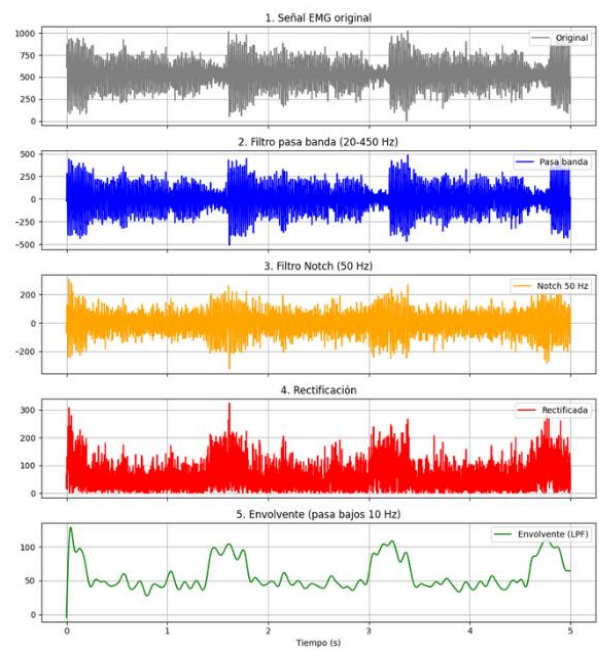

**Figura 6.** Etapas de filtrado de la señal.

### 2.3.2 Generación de Señales EMG Sintéticas

Para mejorar el proceso de entrenamiento y aumentar la cantidad de datos disponibles, se utiliza una red neuronal U-Net para generar señales EMG sintéticas [12]. En este caso se aplica para la generación de datos sintéticos. Esta red toma las señales EMG preprocesadas y aprende a generar nuevas señales que son similares a las originales.

La generación de datos sintéticos resulta especialmente útil cuando se cuenta con un conjunto de datos limitado, como en el caso de las señales EMG, donde obtener grandes cantidades de datos de calidad puede ser costoso o difícil.

### 2.3.3 Modelos Predictivos para Mejoría y Autocalibración

Una vez entrenada la red U-Net para generar señales EMG sintéticas, el siguiente paso propuesto es aplicar inteligencia artificial para la auto calibración del umbral de activación muscular mínimo para considerar satisfactorio un ejercicio. De esta manera, el sistema puede ofrecer un seguimiento más preciso de la evolución del paciente, ayudando en la personalización de los tratamientos y optimizando la eficacia de la rehabilitación.

## 3. Resultados

EquiSay demuestra ser funcional y eficaz en el proceso de rehabilitación del pie equino ya que permite una mejora en la dorsiflexión durante la marcha y una mayor estabilidad del paciente. Además gracias a su diseño ligero, cómodo y ajustable facilita su integración en sesiones de fisioterapia y rehabilitación de pacientes con pie equino sin interferir en la movilidad natural.

Proporciona una base de datos útil que permite realizar una valoración objetiva de la evolución del paciente al clínico que le trata.

Cabe destacar que aunque el sistema presenta buena respuesta por su esquema de procesamiento de la señal de EMG recibida, dado la complejidad de detección de este tipo de señales, se ha identificado la presencia de ciertos artefactos que pueden afectar la precisión de la señal, por lo que se plantea mejorar el sistema de sensores EMG en futuras versiones del prototipo.

En conjunto, los resultados confirman el potencial de EquiSay como una herramienta útil en la rehabilitación de pacientes con esta patología y la posibilidad de su mejora empleando herramientas como la inteligencia artificial.

## 4. Referencias